\documentclass[prb,superscriptaddress,amsfonts,twocolumn,showpacs,floatfix]{revtex4-1}
\usepackage{amsmath}
\usepackage{amsfonts}
\usepackage{amssymb}
\usepackage{bm}
\usepackage{tabularx}
\usepackage{graphicx,color}

\newcommand{\SRO}{Sr$_2$RuO$_4$}
\newcommand{\Tc}{T_\mathrm{c}}
\newcommand{\Ts}{T_\mathrm{c,SRO}}
\newcommand{\Ic}{I_\mathrm{c}}
\newcommand{\dVdI}{\mathrm{d}V/\mathrm{d}I}

\begin{document}

\title{Topological competition of superconductivity in Pb/Ru/$\bm{\mathrm{Sr_2RuO_4}}$ ~junctions}
\author{Taketomo Nakamura}
\affiliation{Department of Physics, Graduate School of Science, Kyoto University, Kyoto 606-8502, Japan}
\author{R. Nakagawa}
\affiliation{Department of Physics, Graduate School of Science, Kyoto University, Kyoto 606-8502, Japan}
\author{T. Yamagishi}
\affiliation{Department of Physics, Graduate School of Science, Kyoto University, Kyoto 606-8502, Japan}
\author{T. Terashima}
\affiliation{Research Center for Low Temperature and Materials Sciences, Kyoto University, Kyoto 606-8501, Japan}
\author{S. Yonezawa}
\affiliation{Department of Physics, Graduate School of Science, Kyoto University, Kyoto 606-8502, Japan}
\author{M. Sigrist}
\affiliation{Theoretische Physik, ETH Z\"{u}rich, Z\"{u}rich CH-8093, Switzerland}
\author{Y. Maeno}
\affiliation{Department of Physics, Graduate School of Science, Kyoto University, Kyoto 606-8502, Japan}

\date{\today}
\begin{abstract}
We devise a new proximity junction configuration 
where an $s$-wave superconductivity and the superconductivity of \SRO\ interfere with each other.
We reproducibly observe in such a Pb/Ru/\SRO\ junction with a single Pb electrode 
that the critical current $\Ic$ drops sharply just below the bulk $\Tc$ of \SRO\ 
and furthermore increases again below a certain temperature below $\Tc$.
In order to explain this extraordinary temperature dependence of $\Ic$, 
we propose a competition effect involving topologically distinct superconducting
phases around Ru inclusions.
Thus, such a device may be called a ``topological superconducting junction''.
\end{abstract}

\maketitle

Despite extensive studies during the past several decades
the realization of spin-triplet superconductivity has not been thoroughly established.
Among a number of the candidates, the layered superconductor \SRO~\cite{MaenoY1998NAT} 
is a most promising case for spin-triplet pairing.\cite{Leggett_book}
The invariant spin susceptibility across the superconducting transition,
observed by both NMR and polarized neutron scattering, provides indeed strong evidence 
for equal-spin pairing.\cite{IshidaK1998NAT,DuffyJA2000PRL}
The odd parity nature of the orbital wave function, required for triplet states,
is compatible with observation using the $\pi$-junction SQUID experiment.\cite{NelsonKD2004}
There has been accumulated evidence to support 
the spin-triplet scenario.~\cite{MackenzieAP2003RMP,IshidaK2008JPCS,NomuraT2002JPSJ-1}
For the complete confirmation of the superconducting parity of \SRO, however,
it is highly desirable to conduct alternative direct experiments for the parity determination.

An important experiment aiming to detect the parity of \SRO\ was introduced in 1999.~\cite{JinR1999PRB}
It was revealed that the critical current $\Ic$ of Pb/\SRO/Pb junctions with an $s$-wave superconductor Pb is sharply suppressed just below the superconducting transition temperature of \SRO, $\Ts = 1.5$ K, and increases again at lower temperatures.
The anomalous temperature dependence of $\Ic$ was interpreted as due to the change of \textit{the phase difference between the TWO Pb electrodes} from 0 to $\pi$, driven by the odd parity superconductivity of \SRO.~\cite{HonerkampC1998PTP,YamashiroM1999PC}
In this study, we report similar behavior but using Pb/Ru/\SRO\ junctions with essentially different configuration containing only a single Pb electrode.
Thus we newly interpret the anomalous $\Ic$ as due to the change of \textit{the winding number between Pb and \SRO}.
Such behavior has never been reported in other superconductor/normal-metal/superconductor (SNS) junctions with an even-parity spin-singlet superconductor or those with odd-parity candidates UPt$_3$ or UBe$_{13}$.\cite{HanS1986PRL,SumiyamaA1998PRL}
The present finding provides evidence for a most intriguing interplay between the superconductors Pb and \SRO\ connected via Ru metal inclusions, reflecting the distinct pairing symmetry of the two systems.

\begin{figure}[!t]
  \centering
  \includegraphics[width=3in]{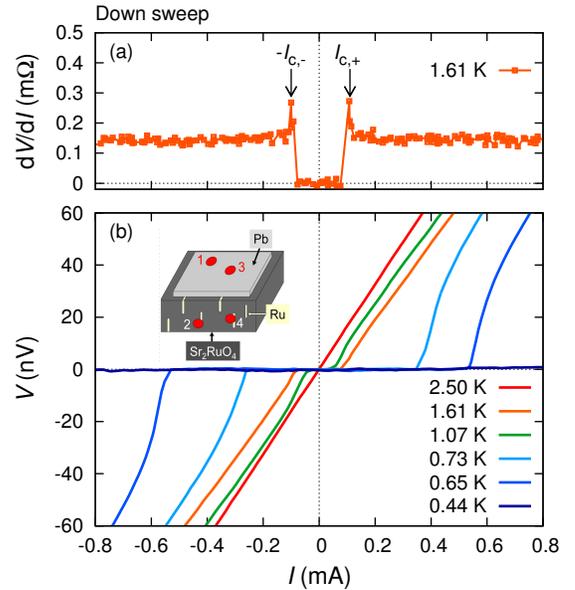}
  \caption[sc]{(color online) 
               (a) Differential resistance vs. current, and 
               (b) voltage-current characteristics of a Pb/Ru/\SRO\ junction.
               The inset to (b) is a schematic of a junction.
               The critical current $I_{\mathrm{c},+}$ is defined at the peak with positive bias current;
               $-I_{\mathrm{c},-}$ with negative bias current.}
  \label{fig:geometry_V-I_dVdI-I}
\end{figure}

An SNS junction provides an opportunity
to examine the quantum interference involving \SRO,
since such interference sensitively affects its voltage-current ($V$-$I$) characteristics.
Technically, poor electrical contact to the surface parallel to the basal $ab$-plane of a \SRO\ crystal 
hampers the fabrication of a high-quality normal-metal/\SRO\ junction.
A Ru metal inclusion in a eutectic \SRO-Ru crystal~\cite{MaenoY1998PRL} 
serves as an ideal normal-metal electrode,
because of its atomically regular interface.\cite{YingYA2009PRL}
This eutectic system exhibits higher onset $\Tc$ of up to 3.5 K
than $\Ts$ ($=1.5$ K) and $\Tc$ of Ru ($\sim 0.5$ K),
thus called the 3-K phase.\cite{KittakaS2009JPSJ}
A number of experiments have revealed that 
the enhanced superconductivity occurs in the \SRO\ region around Ru lamellae
and possesses the same parity as that of \SRO.\cite{MaoZQ2001PRL, SigristM2001JPSJ}

The eutectic crystals were grown by a floating-zone method.\cite{MaoZQ2000MRB}
The $ab$-surface of the crystal cut into the dimension of $\sim 1.9 \times 2.5 \times 0.2~\mathrm{mm}^3$
was polished with diamond slurry.
The area of Ru inclusions at the surface is typically less than 1\% of the total sample area.
A 1-$\mathrm{\mu m}$ thick film was deposited onto the polished surface using 6N-pure lead. 
In order to obtain the $V$-$I$ characteristics between Pb and \SRO\ using a four-probe method,
two terminals each were put on Pb and \SRO\ as illustrated in the inset to Fig.~\ref{fig:geometry_V-I_dVdI-I}(b).

\begin{figure}[!t]
  \centering
  \includegraphics[width=3in]{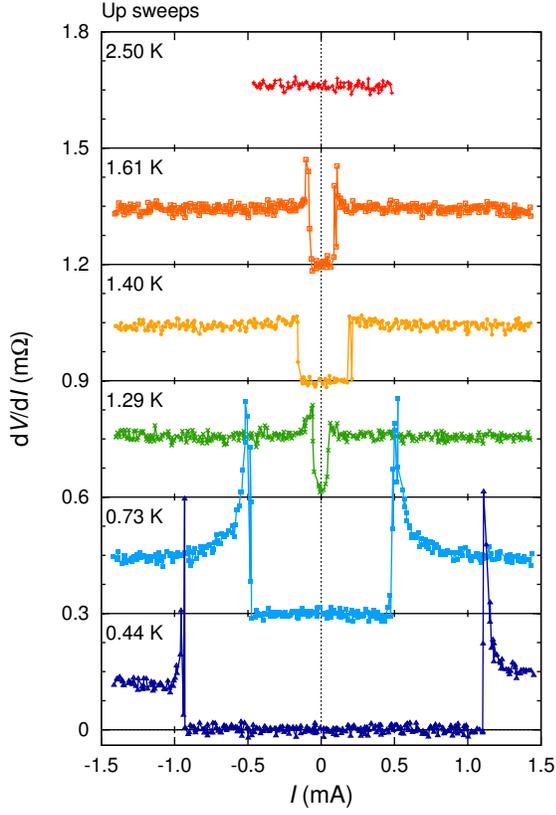}
  \caption[dVdI]{(color online) 
Differential resistance $\dVdI$ of a Pb/Ru/\SRO\ junction at various temperatures.
Each curve is shifted by 0.3 $\mathrm{m}\Omega$ for clarity.
The critical current drops sharply below $\Ts$.}
  \label{fig:dVdI-I_at_variousT}
\end{figure}

The differential resistance $\dVdI$ of Pb/Ru/\SRO\ junctions was measured
by applying an AC current at a frequency of 887 Hz using a lock-in amplifier.
Fig.~\ref{fig:geometry_V-I_dVdI-I}(a) is a representative curve 
of its dependence on the DC bias current at 1.61 K.
Fig.~\ref{fig:geometry_V-I_dVdI-I}(b) represents $V$-$I$ characteristics 
at various temperatures obtained by integrating $\dVdI$,
indicating the behavior of a typical superconducting junction.
At each temperature, the data were taken after the junction was heated to 1.7 K 
and cooled to the measurement temperature with a negative DC current exceeding $-\Ic$.
At the target temperatures, the bias current was 
swept to a positive value (an up sweep) followed by a down sweep.
All the curves in Fig.~\ref{fig:geometry_V-I_dVdI-I} represent the data taken on down sweeps.
Fig.~\ref{fig:dVdI-I_at_variousT} represents the behavior in the up sweeps.
As in Fig.~\ref{fig:geometry_V-I_dVdI-I}(a), 
finite junction voltage emerges with a sharp peak in the differential resistance.
Considering the amplitude of the AC current of 20 $\mathrm{\mu A}$-rms,
the critical current $\Ic$ is defined more precisely and accurately at the peak of $\dVdI$,
rather than at the onset of non-zero $\dVdI$.

\begin{figure}[!t]
  \centering
\includegraphics[width=3in]{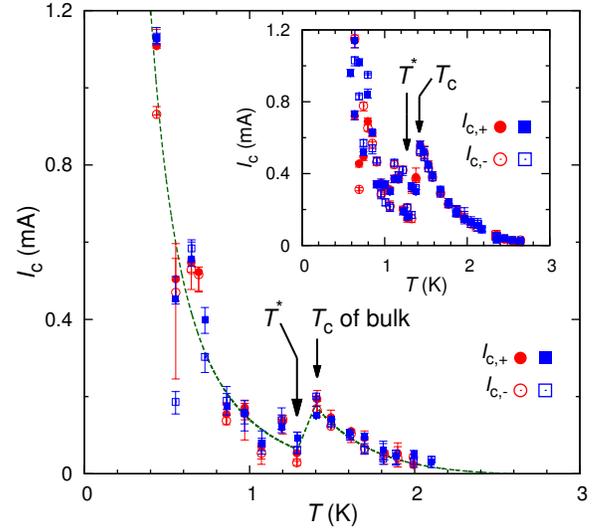}
  \caption[\Ic]{(color online) 
Variation of the critical current $\Ic$ with temperature of a Pb/Ru/\SRO\ junction.
The inset displays $\Ic$ of another Pb/Ru/\SRO\ junction.
The circles and the squares indicate critical currents of up sweeps and down sweeps, respectively. 
$\Ic$ sharply drops just below $\Ts$ but increases again 
below a certain temperature designated as $T^*$.
The broken curve is a guide to the eyes.}
  \label{fig:Ic-T}
\end{figure}

As shown in the top of Fig.~\ref{fig:dVdI-I_at_variousT},
$\dVdI$-$I$ curves are almost flat above 2.1 K.
Below 2.1 K, the curves exhibit a dip-and-peak structure centered at the zero-bias current
and the dip reaches zero at about 1.9 K, which is clearly higher than $\Ts$. 
$\Ic$ increases upon cooling below 2.1 K 
and the $\Ic$-$T$ curve in Fig.~\ref{fig:Ic-T} exhibits a super-linear temperature dependence 
down to $\Ts$:
$\Ic \propto (\Tc-T)^{n}$ with $n=2.6$ for $\Tc=2.5$ K, or $n=3.7$ for $\Tc=3.0$ K.
For a tunneling junction with an insulator between the two superconductors, 
the Ambegaokar-Baratoff (A-B) theory gives 
$\Ic = (eR)^{-1} \Delta_1 K([1-{\Delta_1}^2/{\Delta_2}^2]^{1/2})$
where $R$ is the junction resistance in the normal state, 
$\Delta_{1,2}~(\Delta_{1}<\Delta_{2})$ is the gap function of each superconductor, 
and $K(x)$ is a complete elliptic integral of the first kind.\cite{AmbegaokarV1963PRL}
For $\Delta_1$ substantially smaller than $\Delta_2$, as in the junctions studied here,
the temperature dependence of $\Ic$ should actually be sub-linear, 
contrary to the behavior shown in Fig.~\ref{fig:Ic-T}.
Furthermore, the $\Ic R$ values of our junctions, for example 0.01 $\mu$V at 1.4 K, 
are much smaller compared to an estimate of 400 $\mu$V 
using the A-B theory with $T_\mathrm{c1}=3$ K and $T_\mathrm{c2}=7$ K.
These facts suggest that the Pb/Ru/3-K-phase configurations should be interpreted 
in terms of an SNS device with clean SN interfaces where the proximity effect induced 
coupling plays a crucial role.\cite{FurusakiA1991PRB}

Just below $\Ts = 1.4$ K, used in the device represented here,
the $\Ic$ suddenly drops to nearly zero.
This is a surprising result because 
below this temperature the interfacial superconductivity of the 3-K phase 
develops into the bulk superconductivity of \SRO\ 
and under usual circumstances $\Ic$ is expected to increase rapidly.
This distinct behavior marks clear evidence for unconventional interference effects.
In addition, it indicates a qualitative change of the interfacial state of \SRO\ at $\Ts$.
Below $\Ts$, $\Ic$ remains suppressed in a narrow temperature range
and exhibits complicated temperature dependence.
As additional peculiar behavior, $\Ic$ starts to increase sharply 
below a certain temperature at about 1 K, which we designate as $T^*$.
This anomalous overall temperature dependence is well reproducible among several samples.
Another unusual feature of this junction is that 
the differential resistance as well as the associated $\Ic$ 
exhibit an unusual random variations below $\Ts$,
as represented by the data points in Fig.~\ref{fig:Ic-T}.

\begin{figure}[!t]
  \centering
\includegraphics[width=2.5in]{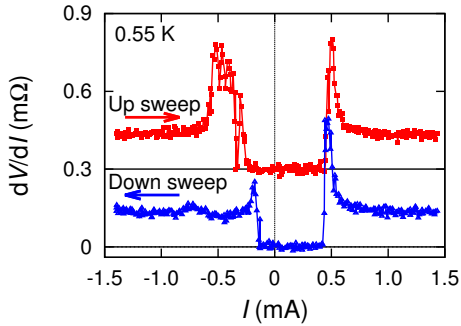}
  \caption[fig4]{(color online). 
Asymmetric and hysteretic bias current dependence of differential resistance $\dVdI$ of a Pb/Ru/\SRO\ junction at 0.55 K.
The up sweep curve is shifted by 0.3 $\mathrm{m}\Omega$.
The down sweep curve was obtained immediately following the up sweep to 1.45 mA.}
  \label{fig:asym}
\end{figure}

In addition to the large variations in $\Ic$, 
$\dVdI$ often becomes asymmetric with respect to the sign of the bias current below $T^*$
as in Fig.~\ref{fig:asym}.
This asymmetric behavior implies the availability of many metastable order parameter configurations,
similar to variable arrangements of the chiral $p$-wave domains in \SRO.\cite{KidwingiraF2006,BouhonA2010NPJ,KambaraH2008PRL}

Previous experiments and theories suggest that the 3-K phase originates most likely from  
the nucleation of superconductivity at the interface of \SRO\ and Ru.\cite{MackenzieAP2003RMP} 
It corresponds to a single $p$-wave component existing in a narrow spatial range on the \SRO\ side with 
its momentum parallel to the interface, denoted as $p_\parallel$-wave.\cite{SigristM2001JPSJ}
This time-reversal symmetry conserving state, called ''A-phase'', corresponds topologically
to a superconducting state without phase winding ($N=0$) 
from the viewpoint of the Ru-inclusion.\cite{KaneyasuH2010JPSJ2} 
Below $ T_{\varepsilon} = 2.4 - 2.6 $ K the time-reversal symmetry breaking appears 
by adding an additional order parameter component, $p_\parallel + i \varepsilon p_\perp$,\cite{SigristM2001JPSJ} 
which may correspond to the ''A'-phase'' with $ N=0 $
or the ''B-phase'' with $ N=1 $ within the theory introduced by Kaneyasu \textit{et al}.\cite{KaneyasuH2010JPSJ2}
The latter is topologically compatible with the chiral $p$-wave bulk phase.

\begin{figure}[!t]
  \centering
\includegraphics[width=3.4in]{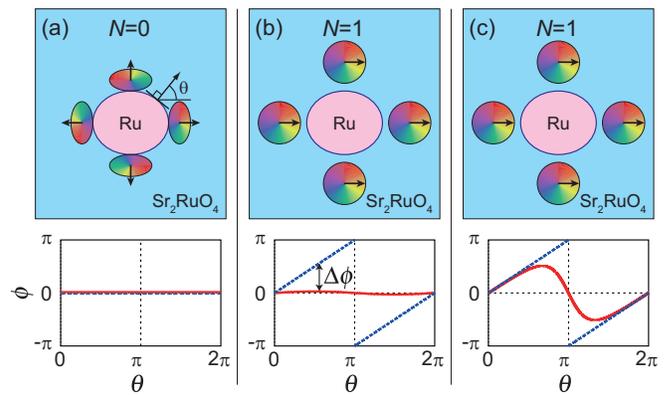}
  \caption[3-K]{(color online). 
Schematic images of the evolution of the order parameter at the \SRO/Ru interface.
In the upper panels, the colors depict 
the momentum-directional dependence of the superconducting phase $\phi$ at each spatial position; 
the arrows denote the momentum direction for which $\phi=0$.
The angle $\theta$ is defined as normal to the interface (see (a)).
The lower panels represent the superconducting phases $\phi(\theta)$ at the \SRO/Ru interface 
under no external current.
The solid lines represent the phase of $s$-wave superconductivity in Ru 
and the broken lines that of $p$-wave superconductivity in \SRO.
(a) $\Ts < T < T_\epsilon$: 
the $p_\parallel \pm i \epsilon p_\perp$ state with the winding number $N=0$ is realized (A'-phase),
matching with the $s$-wave order parameter induced in Ru.
(b) $T \lesssim \Ts$: replacement by 
 $p_x \pm i p_y$, the bulk state of \SRO, with $N=\pm 1$ (B-phase).
(c) $T \ll \Ts$: 
increasing interfacial energy enlarges the phase deformation in the \textit{s}-wave,\cite{KaneyasuH2010JPSJ} 
strengthening the Josephson coupling. 
}
  \label{fig:explanation}
\end{figure}

For the present junction geometry 
we assume that Pb induces superconductivity of $s$-wave symmetry in the Ru inclusions by proximity effect, 
which through spin-orbit interaction yields a direct coupling to the $p_\parallel$-wave order parameter in \SRO.\cite{GeshkenbeinVB1987PRB,SigristM1991RMP}
The topological matching with $N=0$ of Ru naturally favors the A'-phase over the B-phase 
as depicted in Fig.~\ref{fig:explanation}(a), 
because the phase difference $\Delta \phi(\theta)$ 
at the interface can be set to zero for all angles $\theta$ around the circumference 
(junction ground state). 
While the A-phase consisting of only the $p_{\parallel}$-component is strongly localized at the interface, 
the A'-phase with the additional $p_{\perp}$-component is more extended 
(large normal-metal coherence length $\xi_\mathrm{n}$ perpendicular to the interface). 
This extension can explain why $\Ic(T)$ starts to increase slowly 
with lowering $T$ below $T_{\varepsilon}$: 
the gradually growing order parameter $p_{\perp}$ strengthens the superconducting connections 
between different Ru-inclusions and to the external contacts. 
Thus, $\Ic$ is expected to grow approximately as $ \exp[-d/\xi_\mathrm{n}(T)] $ 
with $d$ the average connecting distance among Ru-inclusions and contacts 
and $ \xi_\mathrm{n}(T) \propto (T - \Ts)^{-1/2} $.\cite{KittakaS2009JPSJ}
This yields behavior qualitatively compatible with the experimental results.

With the onset of bulk superconductivity at $\Ts$ 
the order parameter at the interface changes its topology 
to that of the B-phase (Fig.~\ref{fig:explanation}(b)), 
which due to its winding number $ N= \pm 1 $ is frustrated 
with the phase of the $s$-wave order parameter in Ru. 
For the overall Josephson current 
$I = \int_0^{2\pi} d\theta J_\mathrm{c}(\theta,T) \sin \Delta\phi(\theta) 
\approx \bar{J}_\mathrm{c}(T) \int_0^{2\pi} d\theta \sin \Delta \phi(\theta)$,
this topological mismatch leads to an almost complete cancellation.
At the same time,
this frustration induces mild phase deformation just below $\Ts$.
With decreasing temperature, 
the growing interfacial superconductivity in \SRO\ requires less phase mismatch
to minimize the total junction energy.
As a result the region of significant $\Delta\phi$ is confined 
into a shrinking range of $\theta$ (Fig.~\ref{fig:explanation}(c)).\cite{KaneyasuH2010JPSJ}
With the application of external current, 
the resulting phase deformation is such that 
the accompanying Josephson current density $J_\mathrm{c}(\theta, T) \sin \Delta\phi(\theta)$ 
is constructively added and grows at lower temperatures.\cite{KaneyasuH2011inPre}

The extraordinary temperature dependence of $\Ic$ is explained 
in terms of a junction consisting of a topological superconductor 
encapsuling a conventional superconductor.
For this reason, 
such a device may be called a ``topological superconducting junction'', 
in which nontrivial character of a topological superconductor 
becomes observable by appropriate design of the geometry.

The low-temperature phase ($ T < \Ts $) introduces new features. 
One is the variation of the $\dVdI$-vs-$I$ curves 
on the sweep direction of the supercurrent (see Fig.~\ref{fig:asym}). 
This may involve complex phase frustration effects introduced below $ \Ts $ 
through the phase winding of the B-phase, 
which can lead to various metastable states. 
Similar variations have been reported in different junctions of \SRO.\cite{KidwingiraF2006,KambaraH2008PRL}
Since our present junctions contain a number of Ru inclusions acting as parallel contacts,
future experiments using junctions consisting of a single Ru inclusion may resolve this issue.

The coupling of a Pb to \SRO\ via a Ru-inclusion provides a unique opportunity to 
investigate quantum interference effects between topologically incompatible superconducting phases.
The unusual sharp drop of the critical current $\Ic$ below $\Ts$ 
and its curious recovery below $T^*$ can be explained 
by the change of the topology of the superconducting order parameter in \SRO\ 
surrounding the Ru-inclusion.
Thus, the device investigated here can be classified as a new class of superconducting junctions: 
a ``topological superconducting junction''.

We thank Y. A. Ying, S. Kittaka, K. Ishida, T. Sumi and Y. Yamaoka for their support, 
and H. Kaneyasu and D. F. Agterberg for useful discussions.
This work is supported by Grants-in-Aid 
for Global COE program
``the Next Generation of Physics Spun from Universality and Emergence'', 
for Scientific Research on Priority Areas 
``Physics of New Quantum Phases in Superclean Materials''
and 
for Scientific Research in Innovative Areas
``Topological Quantum Phenomena''
from MEXT of Japan.
T. N. is supported by the JSPS.

\end{document}